# Single Rh adatoms stabilized on α-Fe$_2$O$_3$(1$\bar{1}$02) by co-adsorbed water

*Florian Kraushofer[1], Lena Haager[1], Moritz Eder[2], Ali Rafsanjani-Abbasi[1], Zdeněk Jakub[1], Giada Franceschi[1], Michele Riva[1], Matthias Meier[1], Michael Schmid[1], Ulrike Diebold[1], Gareth S. Parkinson[1]*

[1] Institute of Applied Physics, TU Wien, Wiedner Hauptstraße 8-10/E134, 1040 Wien, Austria

[2] Chair of Physical Chemistry & Catalysis Research Center, Technical University of Munich, Lichtenbergstraße 4, 85748 Garching, Germany

**Corresponding Author**

* parkinson@iap.tuwien.ac.at

ABSTRACT: Oxide-supported single-atom catalysts are commonly modelled as a metal atom substituting surface cation sites in a low-index surface. Adatoms with dangling bonds will inevitably coordinate molecules from the gas phase, and adsorbates such as water can affect both stability and catalytic activity. Here, we use scanning tunneling microscopy (STM), noncontact atomic force microscopy (ncAFM), and x-ray photoelectron spectroscopy (XPS) to show that high densities of single Rh adatoms are stabilized on α-Fe$_2$O$_3$(1$\bar{1}$02) in the presence of $2 \times 10^{-8}$ mbar of water at room temperature, in marked contrast to the rapid sintering observed under UHV conditions. Annealing to 50 °C in UHV desorbs all water from the substrate leaving only the OH groups coordinated to Rh, and high-resolution ncAFM images provide a direct view into the internal structure. We provide direct evidence of the importance of OH ligands in the stability of single atoms, and argue that their presence should be assumed when modelling SAC systems.

**TOC GRAPHICS**

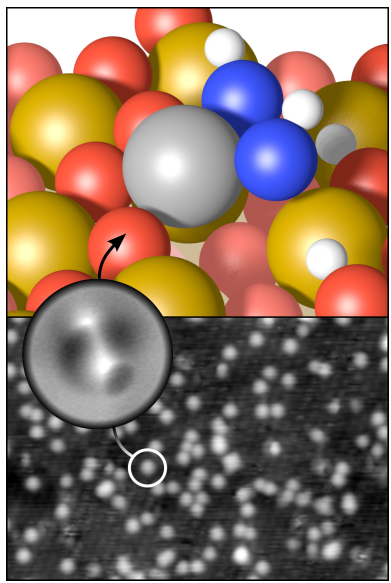

Understanding how metals bind to oxide supports has long been a goal of catalysis research. The issue is particularly pressing in the field of single-atom catalysis (SAC), because the coordination environment of the active site not only decides whether the metal atoms will be stable against thermal sintering, but also strongly affects the catalytic properties. Since transmission electron microscopy (TEM) studies usually assign the isolated adatoms to be located in cation-like sites relative to the bulk structure,[1-7] bulk-continuation or substitutional sites are commonly used as a starting point for exploring reaction pathways.[1, 8-11] A more realistic model would account for the extensive hydroxylation of the support that usually occurs in reactive environments, but given the difficulty ascertaining reliable information on the local structure from experiment, this additional complexity is usually omitted. It can be important, however, as coordinating additional OH to Rh adatoms has been shown to strongly affect the binding energy of CO at that site,[12] and adsorbed water can play a direct role in catalytic mechanisms.[13-14]

Here, we explore the stability of single Rh adatoms on hematite ($\alpha$-Fe$_2$O$_3$), with and without the presence of water. While the $\alpha$-Fe$_2$O$_3$(0001) surface is commonly used to represent FeO$_x$ catalysts in computations,[1, 8-11] it is a poor choice for a model system because the atomic-scale structure is unclear even under UHV conditions.[15-16] We instead utilize the ($1\bar{1}02$) facet, on which a monophase bulk-truncated ($1 \times 1$) termination can reproducibly be prepared.[17-20] Moreover, the surface structure and its interaction with water are well understood.[17-18] The surface termination[17] and room temperature water adsorption sites[18] are briefly introduced in Figure S1. We show that isolated Rh atoms form only in the presence of water vapor; rapid sintering occurs already at room temperature in UHV conditions. The stabilization occurs through the coordination of multiple OH ligands, which remain on the surface at 50 °C after all other water has desorbed.

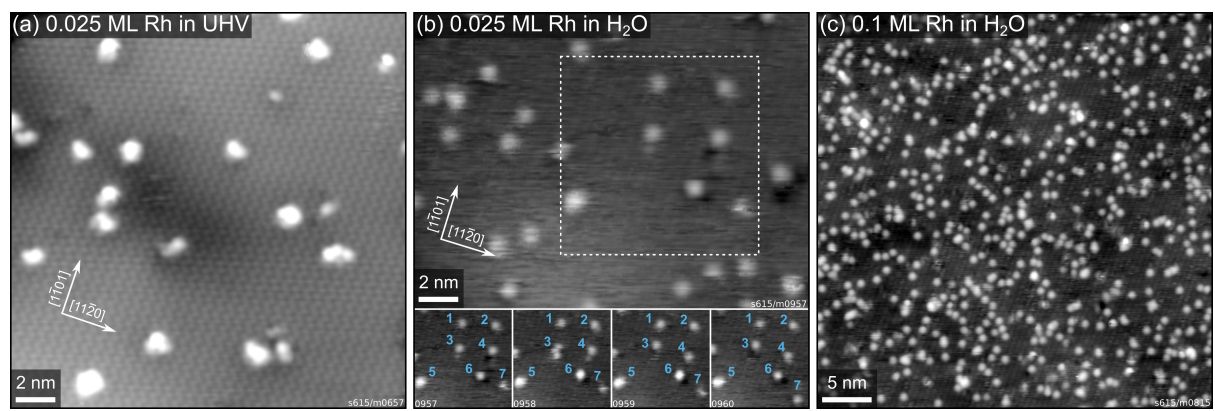

**Figure 1.** Rh stabilization by $H_2O$ on α-$Fe_2O_3$(1$\bar{1}$02). (a) STM image ($U_{sample}$ = +3 V, $I_{tunnel}$ = 0.3 nA) of 0.025 ML Rh on α-$Fe_2O_3$(1$\bar{1}$02), deposited at room temperature in UHV. (b) STM image ($U_{sample}$ = +3 V, $I_{tunnel}$ = 0.2 nA) of 0.025 ML Rh on α-$Fe_2O_3$(1$\bar{1}$02), deposited at room temperature in a background of 2 × $10^{-8}$ mbar $H_2O$. Consecutive STM images from the area indicated by the dashed square are shown in the bottom row, with the same features labelled in blue in each frame. (c) STM image ($U_{sample}$ = +2 V, $I_{tunnel}$ = 0.3 nA) of 0.1 ML Rh on α-$Fe_2O_3$(1$\bar{1}$02), deposited in a partial pressure of 2 × $10^{-8}$ mbar $H_2O$.

Figure 1 shows STM data taken after depositing Rh on the α-$Fe_2O_3$(1$\bar{1}$02) surface in UHV, and in a partial pressure of 2 × $10^{-8}$ mbar $H_2O$. In the case of room temperature deposition in UHV [Figure 1(a)], this surface does not stabilize single Rh adatoms, which instead form small clusters after deposition.[20] In contrast, after depositing with background $H_2O$, the majority of features in Fig. 1(b) and (c) appear as uniform isolated protrusions. To determine how many Rh atoms each feature contains, we performed a separate experiment in which we again deposited Rh in background $H_2O$, evaluated the feature density in STM, then annealed the surface in oxygen at 520 °C. We have shown previously that this results in Rh being incorporated in the first subsurface layer,[20] where it is still imaged as well-defined, isolated features in STM, without significant loss of Rh to the bulk. Because the same density of features was found before and after incorporation (shown in Figure S3 and corresponding text), we assign the bright features in Figure 1(b, c) as single Rh adatoms stabilized by water.

At low Rh coverage, the features are occasionally mobile, as can be seen in consecutive STM images taken from the same area. In the four frames shown in Fig. 1(b), the features labelled as 3, 4 and 7 move over the course of the acquisition, while the rest remain in place. Furthermore, we observe that some of the features are imaged with increased apparent height, but that they can switch between the two apparent heights from one frame to the next. This is the case for the feature labelled as 6 in Figure 1(b), which appears brighter in the second and third frame. The two different apparent heights are measured as $(108 \pm 6)$ pm and $(223 \pm 13)$ pm with respect to the water-covered substrate. Despite the mobility of the features, no agglomeration to clusters was observed. This suggests that the adatoms are stabilized thermodynamically instead of kinetically, i.e., diffusion barriers are low, but it is energetically favourable to keep the atoms separated in the presence of water.

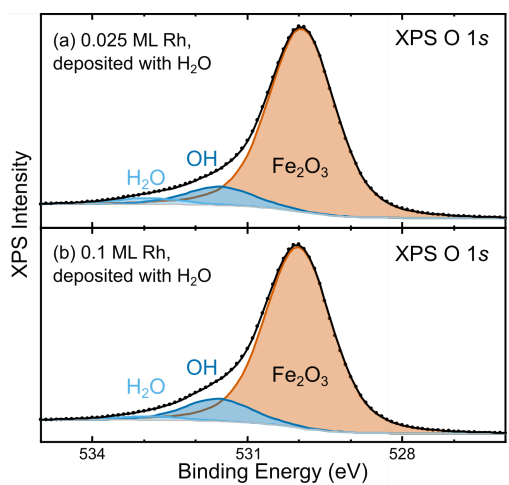

**Figure 2.** The O $1s$ region in XPS (Al Kα, 70° grazing emission, pass energy 16 eV) of Rh stabilized by $H_2O$ on α-$Fe_2O_3$(1$\bar{1}$02). Spectra for (a) 0.025 ML and (b) 0.1 ML Rh deposited in $2 \times 10^{-8}$ mbar $H_2O$ correspond to the STM images in Figure 1(b) and (c). The data (points) were fitted (solid lines) with a component corresponding to lattice $O^{2-}$ anions at 529.9 eV and contributions from molecular $H_2O$ (532.9 eV) and OH (531.5 eV).[18]

XPS data of the O $1s$ region corresponding to the STM images in Figure 1(b, c) are shown in Figure 2. The main contribution at 529.9 eV is assigned to lattice oxygen, and two additional contributions at 532.9 eV and 531.5 eV are assigned to molecular $H_2O$ and OH groups, respectively.[18] It is worth noting that in both cases, the O $1s$ peak contains a significantly higher fraction of OH in the presence of Rh than on the pristine surface. In the absence of rhodium, the $H_2O$/OH ratio is consistently about 0.40 for any submonolayer water coverage.[18] Since dissociation results in two hydroxy groups per $H_2O$, this ratio corresponds to 45% of the water being molecularly adsorbed. Instead, in the presence of Rh, we find an $H_2O$/OH ratio of 0.32 for 0.025 ML Rh, and of only 0.17 for 0.1 ML Rh. In terms of how much water is molecularly adsorbed, this would correspond to only 39% and 26%, respectively. Thus, more water is clearly being dissociated with increasing Rh coverage, which suggests that Rh atoms are active sites for water dissociation. XPS C $1s$ data was also routinely collected of the pristine surface, after Rh deposition and after annealing. We found no sign of contamination within the detection limit of the instrument (<0.01 ML carbon atoms), allowing us to rule out contributions of carbonaceous species to the O $1s$ spectra.

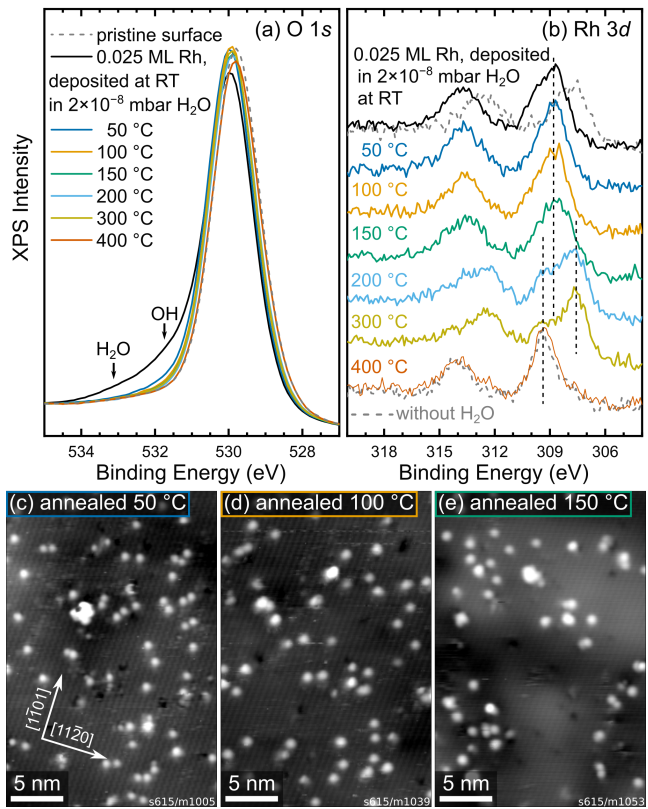

**Figure 3.** Thermal stability of $H_2O$-stabilized Rh on α-$Fe_2O_3$(1$\bar{1}$02). (a, b) XPS data (Al Kα, 70° grazing emission, pass energy 16 eV) of the O $1s$ and Rh $3d$ regions, respectively, of 0.025 ML Rh deposited on α-$Fe_2O_3$(1$\bar{1}$02) at room temperature in a partial pressure of $2 \times 10^{-8}$ mbar $H_2O$, and after successive annealing steps (10 min each) in UHV at different temperatures. The as-deposited spectra (black) correspond to the data shown in Figure 1(b) and Figure 2(a). In panel (b), corresponding spectra for the same Rh coverage, but deposited without $H_2O$, are shown for comparison (grey, dashed). The positions of the initial peak maximum (308.9 eV) and the two main components at elevated temperatures (307.6 eV and 309.3 eV) are marked by the vertical dashed lines. (c-e) STM images taken after the first three annealing steps, corresponding to the blue, orange and green lines in (a) and (b): (c) 50 °C ($U_{sample}$ = +3 V, $I_{tunnel}$ = 0.2 nA), (d) 100 °C ($U_{sample}$ = +3 V, $I_{tunnel}$ = 0.1 nA), (e) 150 °C ($U_{sample}$ = +3 V, $I_{tunnel}$ = 0.1 nA).

To investigate the thermal stability of water-stabilized Rh adatoms, we performed consecutive heating steps after deposition of 0.025 ML Rh in $2 \times 10^{-8}$ mbar water. After each step, the sample was cooled to room temperature to acquire XPS and STM data, shown in Figure 3. The O $1s$ peak [Figure 3(a)] shifts to higher binding energy by 0.2 eV immediately after deposition, most likely due to band bending caused by the adsorbates. The component corresponding to $H_2O$ (532.9 eV) disappears after annealing at 50 °C for 10 min, but a small shoulder corresponding to OH (531.5 eV) remains, accounting for 1.2% of the O $1s$ peak area. Fits to the data are shown in Figure S2. The OH peak area further decreases to 0.5% of the total O $1s$ peak area when heating to 100 °C and to noise level at higher temperatures. The O $1s$ peak then remains unchanged until Rh is fully incorporated at 400 °C, at which point the peak maximum shifts back to match the one of the pristine surface. The Rh $3d$ peak [Figure 3(b)] initially has its maximum at $\approx 308.9$ eV, significantly higher than when Rh is deposited without water (307.8 eV, dashed grey line).[20] The maximum remains at this position while heating to 100 °C and 150 °C. At higher temperatures, the behavior closely resembles that previously observed in the absence of water: A component corresponding to clusters (307.6 eV) starts developing at 150 °C and reaches a maximum at 300 °C, and a component corresponding to incorporated Rh (309.3 eV) first appears at 200 °C, and finally accounts for almost the entire Rh peak at 400 °C.[20]

In STM, most of the single features remain after annealing to 50 °C [Fig. 2(c)], although some clusters are also visible. Interestingly, the single features differ from the state directly after deposition (Figure 1) in that they are immobile and do not exhibit the switching of apparent height observed at room temperature. In the absence of Rh, all water is already desorbed from the support at this temperature,[18] so the remaining OH visible in XPS is likely bound to the Rh adatoms. After heating to 100 °C and 150 °C [Figure 2 (d) and (e)], the number of single features is reduced in each temperature step as Rh sinters to small clusters. Since almost no signature of OH or $H_2O$ remains in XPS after heating to 100 °C, it seems plausible that at this point, sintering is limited purely by diffusion kinetics. Measuring only single features, we also find a slightly lower apparent height in STM after annealing to 100 °C [$(152 \pm 13)$ pm] than after annealing to 50 °C [$(180 \pm 19)$ pm]. Interestingly, while Rh deposited without water at room temperature sinters immediately, some single atoms remain even when all water has been desorbed. This suggests that the temporary coordination to water facilitates a metastable configuration that is not directly accessible at room temperature in UHV.

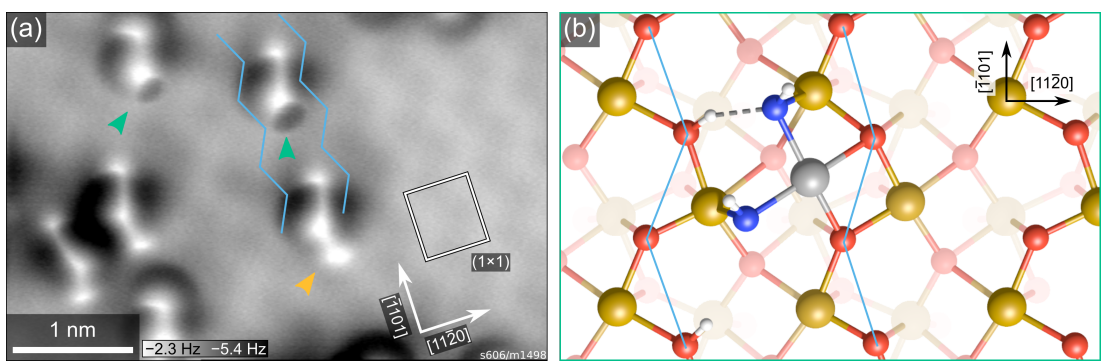

**Figure 4.** Structure of $H_2O$-stabilized Rh on α-$Fe_2O_3$(1$\bar{1}$02). (a) ncAFM image acquired at liquid He temperature of 0.05 ML Rh on α-$Fe_2O_3$(1$\bar{1}$02), deposited at room temperature in a partial pressure of $2 \times 10^{-8}$ mbar $H_2O$, then heated to 80 °C to desorb all water not coordinated to Rh. (b) Schematic model (top view) for the features indicated by green arrows in panel (a). A Rh adatom (grey) is stabilized by two OH groups ($O_{water}$ in blue, hydrogen in white). The zigzag rows of surface oxygen are marked in blue in both panels. The orange arrow highlights a third protrusion that is sometimes present, which we tentatively attribute to an additional water molecule atop a surface Fe and hydrogen bonded to one of the OH groups. Additional side and perspective views of the model in panel (b) are shown in Figure S4.

To further characterize the configuration of water-stabilized Rh on $\alpha$-Fe$_2$O$_3$(1$\bar{1}$02), we performed the same deposition in $2 \times 10^{-8}$ mbar H$_2$O as before in a different UHV chamber, annealed the sample to remove the water not coordinated to Rh, then acquired ncAFM images with a CO-terminated tip at liquid He temperature ($\approx 4$ K). The results are shown in Figure 4(a). Two different motifs are observed, consisting of either two or three bright features on top of a darker area, marked by green and orange arrows in Figure 4(a), respectively. Both motifs also occur in a mirrored form, as expected based on the surface symmetry. In the absence of water, Rh adatoms are imaged as dark features in ncAFM at these scanning conditions, and we therefore assign the bright features to co-adsorbed water. Based on the position of the features with respect to the underlying surface we propose the model for the motif containing two bright features shown in Figure 4(b). Here, Rh is coordinated in a square-planar geometry with two bonds to surface oxygen and two bonds to OH groups, which also bind to surface iron. Importantly, these OH groups sit at the same sites as water adsorbed on the pristine $\alpha$-Fe$_2$O$_3$(1$\bar{1}$02) surface.[18] It seems likely that an additional water molecule can be added at one end of the complex, resulting in an HO-H$_2$O configuration characteristic of water adsorbed on $\alpha$-Fe$_2$O$_3$(1$\bar{1}$02) in the absence of Rh.[18] This would explain the motifs consisting of three bright features, marked by orange arrows in Figure 4(a). In principle, the Rh adatom may also form an additional bond to a lattice oxygen atom below it, as indicated in Figure S4(b). However, since that oxygen atom is already in a bulk-like 4-fold coordination environment, we assume that such a bond would be significantly weaker and would have only minor effects on the overall Rh stability.

The assumption of two OH groups coordinated to each Rh adatom fits with the OH contribution in XPS (Figure S2) if approximately six substrate oxygen layers contribute to the XPS signal. This would correspond to a probing depth of 5.9 Å and an electron path length in the crystal of 17.3 Å at 70° grazing emission, which is on the order of magnitude of the inelastic mean free path, estimated as $\approx 20$ Å at $E_{kin} \approx 950$ eV.[21-22] Therefore, although experimental and theoretical uncertainties do not allow direct determination of the OH coverage from XPS, the proposed model plausibly fits the XPS O 1$s$ data.

Overall, our data show that single Rh adatoms are stabilized on α-Fe$_2$O$_3$(1$\bar{1}$02) in the presence of water. Interestingly, it appears that two different stabilization mechanisms are at work depending on the water coverage. When Rh is deposited, a complete monolayer of water is likely present on the surface. Under these conditions, we find that the adatoms are mobile, but do not agglomerate, indicating thermodynamic stabilization, i.e., single atoms are energetically more favourable than dimers. Using the direct ncAFM imaging of the adsorbed Rh(OH)$_x$ complex (Figure 4) allows us to rationalize this stability through the square-planar coordination of the adatoms, which is expected to be stable by analogy to Rh coordination complexes. This was previously observed for Rh and Ir adatoms on Fe$_3$O$_4$(001).[23-24] Since the two OH ligands are located in the same sites as water adsorbed at room temperature on the pristine α-Fe$_2$O$_3$(1$\bar{1}$02) surface,[18] no rearrangement is required to accommodate ad-Rh species as they are deposited. The facile diffusion of adatoms after deposition may be explained by Rh diffusing underneath the water ad-layer, as two equivalent 4-fold coordinated sites are available in each unit cell. In contrast, once surrounding water is desorbed, diffusion requires displacing the entire Rh(OH)$_x$ complex at once, breaking not only the Rh-O(H) bonds but also the bonds of OH to surface Fe, thus likely resulting in higher diffusion barriers.

The switching of apparent height observed in STM [Figure 1(a)] is most likely due to adsorption and desorption of either molecules from the residual gas, or of molecular H$_2$O from the surface. In contrast to incorporated 6-fold coordinated Rh, which form after annealing in O$_2$,[20] these adatoms thus have the capability to act as reaction centres. In the model shown in Figure 4(b), we speculate that binding another ligand may allow the Rh adatom to switch from a square-planar to an octahedral coordination by forming an additional bond with the surface oxygen atom directly below it (initial distance ≈2.3 Å in the structure as drawn, see Figure S4). If this process is facile, it would be conceptually similar to Wilkinson's catalyst, in which Rh switches between square-planar and octahedral coordination. In general, while single atoms stabilized entirely through coordination to lattice oxygen tend to be catalytically inactive,[25] we can expect stabilization of adatoms through OH groups to allow for more dynamic reaction kinetics since the adatoms are less strongly oxidized and the ligands easier to displace. Moreover, coordination to OH groups likely allows more flexibility in the structure than the rigid oxide lattice.

In addition to stabilizing the adatoms in an active geometry, the presence of water may also directly contribute to reaction pathways, as proposed for low-temperature CO oxidation on the Au/TiO$_2$ system.[14] Even more substantial involvement has been demonstrated for Pt$_1$/CeO$_2$, where water oxygen is used for CO oxidation via an intermediate step in a Mars-van-Krevelen (MvK) process.[13] It seems plausible that similar abstraction of oxygen from OH groups may be relevant more generally, especially when MvK pathways have been proposed to explain oxidation pathways at or near room temperature occurring on substrates with high oxygen vacancy formation energies.

The high XPS core-level binding energy observed here for Rh $3d$ [Figure 3(b)] indicates that the electronic structure of the adatoms is also strongly modified by their coordination to water. Clearly, any change to the Rh oxidation state would also affect its interaction with reactants, affecting both infrared frequencies and reaction barriers. A similar effect was previously proposed for Rh adatoms interacting with OH groups on anatase TiO$_2$.[12] However, this Rh-OH configuration on anatase only occurred after reduction treatment in $H_2$ gas.[12] In contrast, dissociated water is stably adsorbed on the $Fe_2O_3(1\bar{1}02)$ surface at room temperature, and favourable adsorption sites for Rh adatoms are available even without rearranging the water overlayer, and so a Rh-OH configuration is expected to occur in most realistic conditions. Our results therefore strongly suggests that co-adsorption of water must be accounted for in theoretical studies to accurately model reaction pathways.

**Experimental Methods**

Room-temperature STM and XPS results were collected in a UHV setup consisting of a preparation chamber (base pressure $< 10^{-10}$ mbar) and an analysis chamber (base pressure $< 5 \times 10^{-11}$ mbar). This system is equipped with a nonmonochromatic Al K$\alpha$ X-ray source (VG), a SPECS Phoibos 100 analyser for XPS, and an Omicron $\mu$-STM. The STM was operated in constant-current mode using electrochemically etched W tips. STM images were corrected for distortion and creep of the piezo scanner, as described in ref. [26]. Apparent heights of adatom features were measured with respect to the mean height of a ring around the feature of interest. Noncontact AFM results were acquired in a separate vacuum setup using an Omicron LT-STM equipped with a QPlus sensor and an in-vacuum preamplifier.[27] Rh was deposited from a rod with an electron-beam evaporator (Focus), using a quartz-crystal microbalance to calibrate the deposition rate, with deposition times of ca. 30-120 seconds for 0.025-0.1 monolayers (ML) of rhodium. A repelling bias of 1 kV was used during deposition to avoid implantation of Rh ions. Throughout this paper, we define a monolayer as the number of Fe atoms in the surface layer. 1 ML of Rh is therefore defined as two Rh atoms per $\alpha$-$Fe_2O_3(1\bar{1}02)$-$(1 \times 1)$ unit cell, which corresponds to a density of $7.3 \times 10^{14}$ atoms cm$^{-2}$. The evaporation rate was always calibrated in UHV, even when the deposition was carried out in a background of $H_2O$, which results in errors in the actual coverages. The nominal coverages given in this work are not corrected for these errors and probably overestimate the Rh coverage (see Fig. S3). When depositing Rh with a background of water, the sample was always first exposed to 2 L $H_2O$ (1 L = 1.33 × $10^{-6}$ mbar × s) to ensure a constant $H_2O$ coverage during the deposition.

The experiments were conducted on single-crystalline, 0.03 at.% Ti-doped hematite films grown homoepitaxially by pulsed laser deposition on natural α-$Fe_2O_3$(1$\bar{1}$02) samples (SurfaceNet GmbH, 10 × 10 × 0.5 $mm^3$, <0.3° miscut), as described in detail elsewhere.[19-20] This ensures sufficient conductivity of the samples for STM without reducing the oxide. The surface appears identical to the undoped samples studied previously.[17] Before each experiment, the sample was re-prepared by sputtering (1 keV $Ar^+$ ions, ~2 μA, 15 min) and annealing in oxygen (2 × $10^{-6}$ mbar, 520 °C) for 30 min.

ASSOCIATED CONTENT

**Supporting Information**. STM images of Rh deposited in $H_2O$ background before and after incorporation and description of the coverage quantification.

AUTHOR INFORMATION

E-Mail: parkinson@iap.tuwien.ac.at

**Notes**

The authors declare no competing financial interest.

ACKNOWLEDGMENT

GSP, FK, LH, MM, and ARA acknowledge funding from the European Research Council (ERC) under the European Union's Horizon 2020 research and innovation programme (grant agreement No. 864628). ZJ was supported by the Austrian Science Fund (FWF, Y847-N20, START Prize). GF and UD acknowledge funding from the European Research Council (ERC) under the European Union's Horizon 2020 research and innovation programme (grant agreement No. 883395, Advanced Research Grant 'WatFun'). ME acknowledges support from the TUM-GS of the TU Munich.

**Supporting Information**

# Single Rh adatoms stabilized on α-Fe$_2$O$_3$(1$\bar{1}$02) by co-adsorbed water

*Florian Kraushofer[1], Lena Haager[1], Moritz Eder[2], Ali Rafsanjani-Abbasi[1], Zdeněk Jakub[1], Giada Franceschi[1], Michele Riva[1], Matthias Meier[1], Michael Schmid[1], Ulrike Diebold[1], Gareth S. Parkinson[1]*

[1] Institute of Applied Physics, TU Wien, Wiedner Hauptstraße 8-10/E134, 1040 Wien, Austria

[2] Chair of Physical Chemistry & Catalysis Research Center, Technical University of Munich, Lichtenbergstraße 4, 85748 Garching, Germany

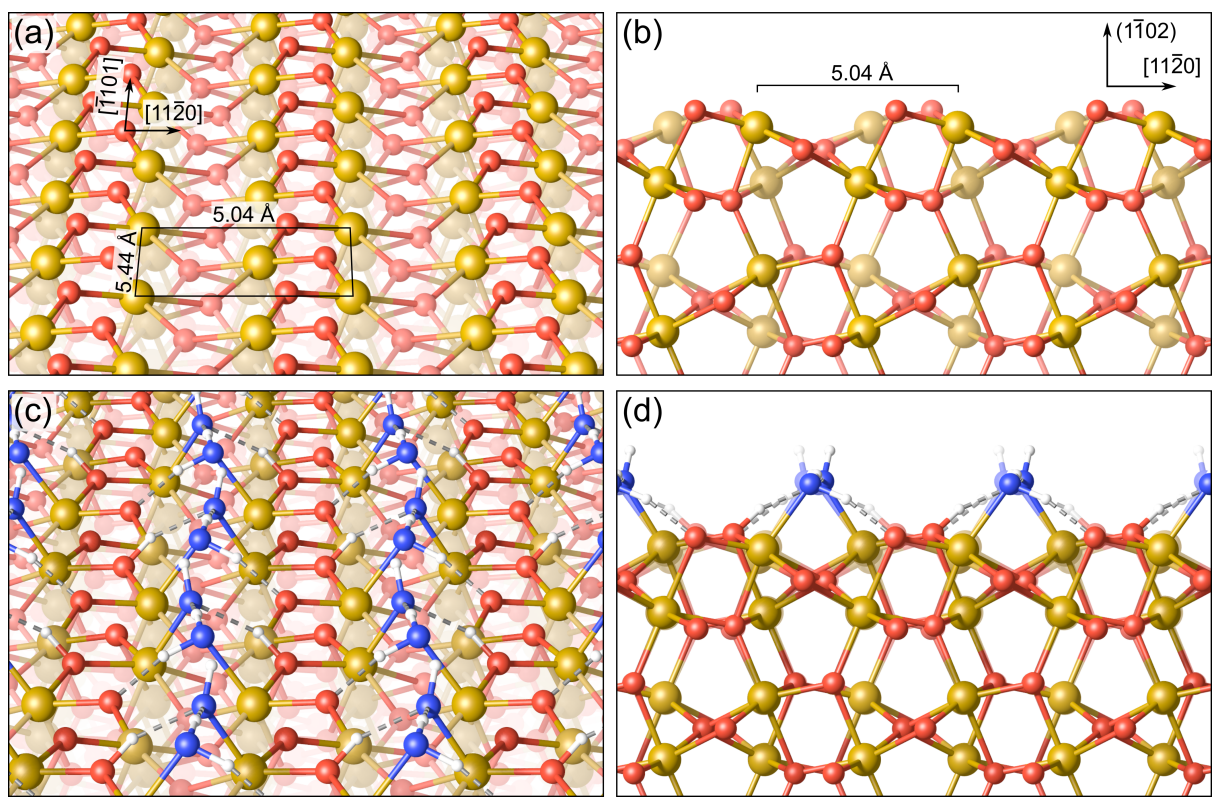

Figure S1: (a, b) perspective and side view of the α-Fe$_2$O$_3$(1$\bar{1}$02)-(1×1) termination.[1] Oxygen and iron are drawn as red (small) and brown (large), respectively. The direction perpendicular to the surface is marked as (1$\bar{1}$02) in round brackets because there is no integer-valued vector in that direction. Oxygen and iron are 4-fold and 6-fold coordinated in the bulk, respectively, and 3-fold and 5-fold coordinated at the surface. (c, d) The same termination with adsorbed water in the configuration before the highest-temperature desorption peak (345 K, 2/3 of a monolayer).[2] Water is adsorbed in the form of half-dissociated dimers. Each water molecule and OH group is coordinated to one surface iron atom, and every third surface iron atom remains unoccupied at this coverage. O$_{water}$ is drawn in blue, hydrogen in white.

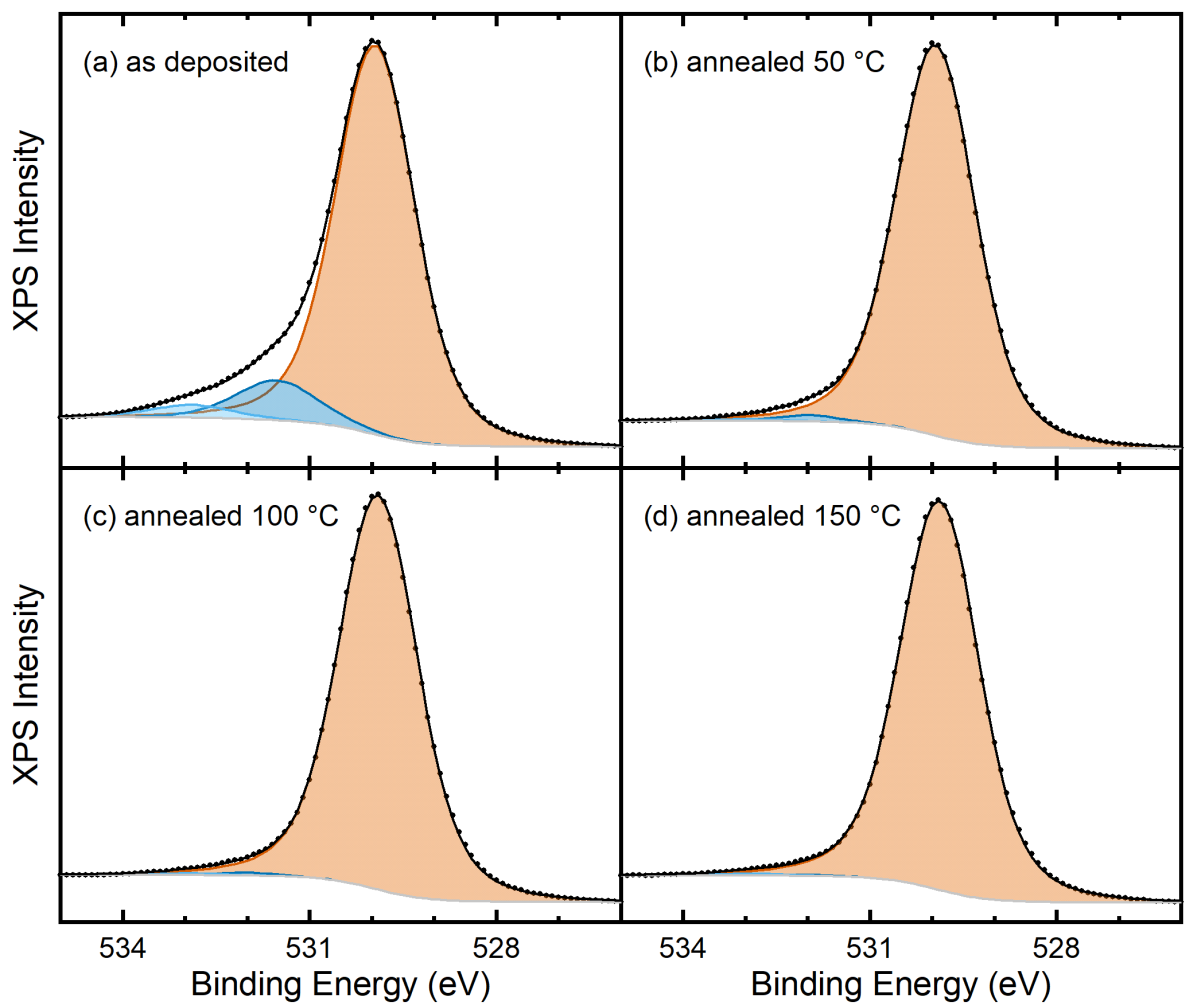

Figure S2: Fits of the O $1s$ region in XPS (Al Kα, 70° grazing emission, pass energy 16 eV) of 0.025 ML Rh stabilized by $H_2O$ on α-$Fe_2O_3$($1\bar{1}02$) before and after annealing, also shown without fits in Figure 3(a). The data (points) were fitted (solid lines) with a component corresponding to lattice $O^{2-}$ anions at 529.9 eV and contributions from molecular $H_2O$ (532.9 eV) and OH (531.5 eV).[2] The substrate peak shape has been optimized to match the pristine sample before Rh deposition. The contributions of the OH components to the overall peak areas are (a) 8.9%, (b) 1.2%, (c) 0.5%, (d) 0.2%. Errors due to data noise are negligible, but systematically larger OH contributions can be obtained if the as-prepared sample is assumed to already be hydroxylated.

## Quantification of deposited rhodium

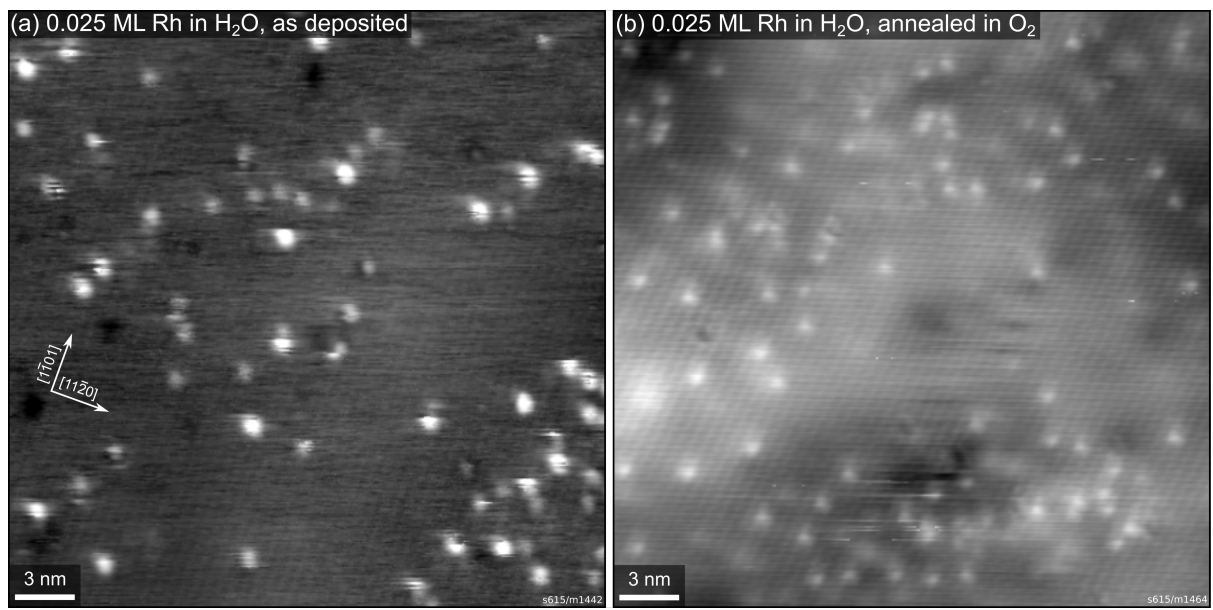

Figure S3: STM images of rhodium on α-Fe$_2$O$_3$(1$\overline{1}$02) (a) after depositing a nominal dose of 0.025 ML Rh at room temperature in a background of $2 \times 10^{-8}$ mbar H$_2$O ($U_{sample}$ = +3.0 V, $I_{tunnel}$ = 0.3 nA), and (b) after annealing the same sample at 520 °C in $2 \times 10^{-6}$ mbar O$_2$ for 30 min ($U_{sample}$ = −2.8 V, $I_{tunnel}$ = 0.1 nA).

To determine whether the bright features found after Rh deposition in a background of water corresponded to single rhodium atoms, we analysed the area density after deposition, then annealed the sample at 520 °C in $2 \times 10^{-6}$ mbar O$_2$ for 30 min to incorporate the Rh atoms. This results in well-defined, single Rh atoms substituting Fe in the first subsurface layer without significant loss of Rh to the bulk,[3] which allows us to easily quantify the coverage through STM analysis. STM images before and after incorporation are shown in Figure S3. The observed area densities are significantly lower than the nominal 0.025 ML (with 1 ML defined as two Rh atoms per unit cell, corresponding to a density of $7.3 \times 10^{14}$ atoms/cm$^{-2}$), probably due to the calibration of the deposition rate with a quartz crystal microbalance being performed in UHV, while the actual deposition was performed in water background. However, the densities are approximately the same before and after incorporating the rhodium into the surface, with 0.014 ML in Figure S3(a) and 0.015 ML in Figure S3(b). Therefore, we conclude that each bright feature in Figure S3(a) contains exactly one Rh atom.

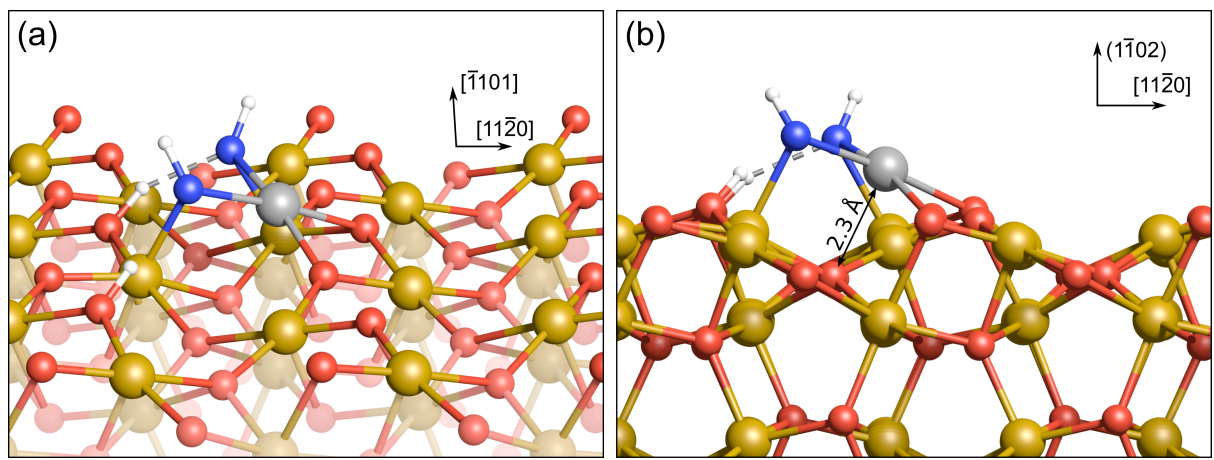

Figure S4: (a) Perspective and (b) side view of the schematic model of a Rh(OH)$_2$ complex shown in Figure 4(b). A Rh adatom (grey) is stabilized by two OH groups (O$_{water}$ in blue, hydrogen in white). The Rh atom may also form an additional bond to a second-layer oxygen atom [distance marked in (b)], but that atom is already fully 4-fold coordinated in a bulk-like site.